# Edge-insensitive Magnetism and Half Metallicity in Graphene Nanoribbons

Shiyuan Gao [1] and Li Yang [1, 2] *

(1) Department of Physics, Washington University in St. Louis, St. Louis, Missouri 63130, USA

(2) Institute of Materials Science and Engineering, Washington University in St. Louis, St. Louis, Missouri 63130, USA

Realizing magnetism in graphene nanostructures is a decade-long challenge. The magnetic edge state and half metallicity in zigzag graphene nanoribbons are particularly promising [Y.-W. Son, et al., Nature **444**, 347 (2006)]. However, its experimental realization has been hindered by the stringent requirement of the mono-hydrogenated zigzag edge. Using first-principle calculations, we predict that free-carrier doping can overcome this challenge and realize ferromagnetism and half-metallicity in narrow graphene nanoribbons of general types of edge structures. This magnetism exists within the density range of gate-doping experiments (~$10^{13}$ cm$^{-2}$) and has large spin polarization energy up to 17 meV per carrier, which induces a Zeeman splitting equivalent to an external magnetic field of a few hundred Tesla. Finally, we trace the formation mechanics of this edge-insensitive magnetism to the quantum confinement of the electronic state near the band edge and reveal the scaling law of magnetism versus the ribbon width. Our findings suggest that combining doping with quantum confinement could be a general tool to realize transition-metal-free magnetism in light-element nanostructures.

**Text**

Ever since its first isolation in 2004 [1], graphene along with its derivative structures has been a long-standing focal point for nanoelectronics research [2, 3]. Particularly, they have many desired properties for spintronics and spin qubit devices, including high mobility, long spin lifetime, and gate-tunable carrier concentration, in addition to an almost vanishing thickness [4, 5, 6]. However, due to the absence of localized *d* or *f* electrons, magnetism does not naturally appear in pristine graphene, and its realization usually relies on specific transition-metal adatoms, defects, or specific edge structures [7]. One of the most attractive candidates for graphene-based spintronic device is the mono-hydrogenated zigzag graphene nanoribbons (commonly referred as ZGNRs), in which graphene is terminated by the zigzag edge on both sides with single hydrogen atom occupying each dangling bond. It was proposed by Son et al. in 2006 [8] with *ab initio* density functional theory (DFT) calculations that this structure becomes half-metallic under a cross-ribbon electric field. This is because the mono-hydrogenated edge produces a localized edge state [9, 10], which leads to a high electronic density-of-state (DOS) and forms a ferromagnetic (FM) ordering along the edges and an antiferromagnetic ordering at the opposite edges. However, the appearance of this magnetic structure is premised upon precisely realizing the mono-hydrogenated edge. Besides, it was shown that this mono-hydrogenated edge is energetically less stable [11, 12, 13, 14, 15] and would give way to other edge structures, such as the mono- and di-hydrogenated armchair (a11 and a22) edges (Figures 1 (a) and (b)) and a reconstructed zigzag edge with one di- and two mono-hydrogenated sites (z211, Figure 1 (c)), under standard conditions in terms of environment hydrogen concentrations [16]. Unfortunately, GNRs formed by those more stable edges are semiconducting and non-magnetic. This is commonly speculated [17, 18] as the reason why experimental evidence of magnetism in ZGNRs has been scarce and indirect [19, 20] for over a decade since the theoretical prediction.

Recently, a few two-dimensional (2D) semiconductors, including GaSe, α-SnO, and InP$_3$, have been proposed as potential FM or multiferroic materials under free-carrier doping [21, 22, 23]. The mechanism behind their magnetic ordering is their unique Mexican-hat-shaped band structure that contributes to a significant peak in DOS. By tuning the Fermi level near this peak via doping, the electron-electron exchange interactions can overcome the kinetic energy cost and allow the doped carriers to form iterant ferromagnetism and half-metallicity. These studies opened a new path for realizing magnetism in low-dimensional structures without involving localized *d* or *f* electrons. One-dimensional (1D) structures like GNRs have intrinsically more divergent van Hove Singularities (vHSs) in their DOS, which give rise to better chances of realizing magnetism via this mechanism. Particularly, because this is essentially an edge-unrelated quantum confinement effect, it may bypass the stringent requirement for edge structures and offers the potential to realize magnetism and half-metallicity in GNRs.

In this Letter, we consider three types of GNRs as shown in Figure 1, which include the two common types of edges, ones along the armchair and zigzag directions, and different types of edge hydrogen passivation as well. These structures are known to be the energetically most stable ones but unfortunately do not exhibit any magnetism intrinsically [16]. Following the convention in Ref. [16], we denote these edge structures as a11, a22 and z211 respectively, where a/z stands for armchair/zigzag edge, and the number denotes how many hydrogen atoms are bonded with each of the consecutive edge carbon atoms within a unit cell. Additionally, we use a number in front to denote the width by the number of C-C dimers or zigzag chains along GNRs. We do not include the mono-hydrogenated ZGNRs because they are intrinsically antiferromagnetic and doping does not essentially change their edge magnetism [24]. Our calculation is based on the *ab initio* pseudopotential projector-augmented wave DFT method [25] as implemented in the Vienna *ab initio* simulation package (VASP) [26]. The exchange-correlation functional uses the generalized gradient approximation with Perdew−Burke−Ernzerhof parametrization (GGA-PBE) [27]. The details of the calculation are included in the Supplementary.

We first focus on the intensively-studied mono-hydrogenated a11-GNRs, which do not exhibit any magnetism before doping. Presented in Figure 2 are the electronic band structure, DOS, and magnetization density of a 4-a11-GNR under a hole doping density of 0.35 hole/nm (corresponding to an approximate planar charge density of $7\times10^{13}$cm$^{-2}$), obtained using a spin-polarized DFT calculation. *Upon doping, the carriers spontaneously polarize into one spin population and form a FM ground state.* As shown in Figures 2 (a) and (b), the bands corresponding to different spins are split, and the Fermi energy only intersects with the band of a single spin, making this system a perfect half metal. The splitting between the opposite-spin bands at the valence band maximum (VBM) is about 85 meV, equivalent to a Zeeman splitting under a huge external magnetic field of 730 T (assuming a spin *g* factor of 2 and no orbital contribution). We also plot the real-space magnetization density (difference between the density of opposite-spin electrons) of the spin-polarized states. As shown in Figure 2 (c), the magnetization density is distributed around the whole GNR, indicating that it is not originated from the edge. Magnetization density for wider a11-GNRs and other types of GNRs are plotted in the Supplementary Figure 1, which further confirms that the magnetization density does not fall off when moving away from the edge.

This doping-induced magnetism and half metallicity are robust for different doping types and densities, suggesting the possibility of bipolar spintronics applications. Figure 3 (a) shows the magnetic moment and spin polarization energy (difference between the FM and paramagnetic ground-state energy) per carrier as a function of the electron or hole doping density for this 4-a11-GNR. As we can see, for both electron and hole dopings, the spins of the free carriers are completely polarized with a saturated magnetic moment of 1 $\mu_B$/carrier, forming a perfect half-metallic state. The half-metallic FM ground state exists for a wide range of doping density up to 0.1 electron/nm for n-doping and 0.4 hole/nm for p-doping. Beyond this range, the magnetic moment rapidly drops to zero, and a paramagnetic ground state is restored.

Spin polarization energy defines the strength of magnetic orders and determines the spin correlation length in 1D and transition temperature in higher dimensions [28]. Although the magnetic momentum is saturated for nearly the entire FM phase, the spin polarization energy per carrier exhibits an inverted parabola shape with a maximum roughly in the middle. For example, for 4-a11-GNR, as shown in Figure 3 (a), the maximum is $\Delta E_{max} = 7$ meV at a hole density of $n_{max} = 0.2$/nm. For electron doping, the value is comparatively smaller, with $\Delta E_{max} = 3$ meV at $n_{max} = 0.07$/nm.

This doping-induced magnetism and half metallicity are universal in narrow GNRs and can be observed in different widths. Figure 3 (b) shows a wider 7-a11-GNR, which also exhibit the FM ground state under both electron and hole doping. The magnetic moment is fully saturated although the spin polarization energy is reduced to around 3~4 meV per carrier. Generally, magnetism becomes weaker with increasing ribbon width, and for a11-GNRs with a width larger than 1.3 nm, the magnetic order is no longer detectable with a doping density resolution of 0.02/nm. This indicates that quantum confinement is the crucial factor to induce the magnetism and the magnetic order is limited within narrow

GNRs. Recent experiments have demonstrated fabrication of high-quality narrow GNRs [29, 30, 6], making our prediction of immediate interests.

Different edge passivations and edge types are known challenges to realize edge magnetism in GNRs. However, this is no longer a barrier to prevent the quantum-confinement induced magnetism from doped narrow GNRs. As shown in Figure 3 (c), the FM ground state is also observed in doped 6-a22-GNR which has a different passivation of the armchair edge. The magnetic moment per carrier is saturated, and the spin polarization energy reaches above 10 meV per carrier for hole doping, even larger than that of narrower 4-a11-GNR. Moreover, Figure 3 (d) shows the FM ground state in the doped 4-z211-GNR, whose edge is energetically more stable than the mono-hydrogenated zigzag edge [16]. Interestingly, it has the largest spin polarization energy among all our studied GNRs, which reaches 17 meV per carrier at an electron doping density of 0.35 e/nm, larger than that of the armchair-edge GNRs with similar width.

In these studied GNRs, the corresponding range of planer doping density is within the $10^{13}$ cm$^{-2}$ range, comparable with those predicted for the magnetic 2D monolayer GaSe [21]. This doping density is accessible with the electrostatic gate doping methods without the need for dopant atoms, which has been commonly used for a wide range of monolayer 2D materials that have similar electron affinity and ionization potential [31, 32]. It is also worth noting that optical doping that creates electron and hole simultaneously [33] could also lead to magnetism in the same way. Additionally, the spin polarization energy in these GNRs can reach a few times higher than that of GaSe (~ 3 meV) and comes close to that of mono-hydrogenated ZGNR [8]. Finally, we have calculated the magnetic properties of narrow GNRs with defective edges. Magnetism and half-metallicity remain (See Figure 2 of supplementary information). Given the widely observed doping in nanostructures, this magnetism may be helpful for understanding a broad range of controversial measurements.

The magnetism observed in doped GNRs above and its evolution with the ribbon width and doping density can be understood from the Stoner theory of iterant magnetism. In the Stoner theory [34], the spin susceptibility of the paramagnetic state is given by $\chi = \frac{N_0}{1-IN_0}$, where $N_0$ is the DOS at the fermi energy and $I$ is the Stoner parameter, decided by the electron exchange and correlation effects. Because the FM instability occurs when $IN_0 >$

1, both an enhanced DOS and a larger $I$ will increase the likelihood of a FM instability. In a 1D system, because the vHS of DOS diverge as $E^{-1/2}$ near the band extrema, the FM instability is much easier to be realized, which drives the formation of the FM states in doped GNRs. For example, for a 4-a11-GNR, the DOS at the Fermi energy is 2.4/eV per carbon atom at a hole density of 0.1 hole/nm, comparable to the 2-3/(eV·atom) DOS of bulk $3d$ transition metals like iron, cobalt and nickel [35].

The Stoner theory can be quantitatively justified by first-principles DFT results. For a single band with effective mass $m$, the Stoner model correspond to a rigid shift in opposite directions for opposite spins near the band edge:

$$\varepsilon_{k\sigma} = \frac{\hbar^2 k^2}{2m} \pm \frac{I}{2}(n_\uparrow - n_\downarrow),$$

where $I$ is the Stoner parameter that implicitly includes the exchange and correlation effects at mean-field level and $n_\uparrow$, $n_\downarrow$ are the density of spin-up and spin-down carriers. The total energy of the electrons (per unit length) is

$$E_{tot}(n_\uparrow, n_\downarrow) = \int_0^{\varepsilon_F^\uparrow} \varepsilon g(\varepsilon) d\varepsilon + \int_0^{\varepsilon_F^\downarrow} \varepsilon g(\varepsilon) d\varepsilon + I n_\uparrow n_\downarrow,$$

where $g(\varepsilon)$ is the DOS of a single-spin band [36], which, after integrating, gives

$$E_{tot}(n_\uparrow, n_\downarrow) = \frac{\hbar^2 \pi^2}{6m}(n_\uparrow^3 + n_\downarrow^3) + I n_\uparrow n_\downarrow.$$

Therefore, the spin polarization energy as a function of the doping density is

$$\Delta E(n) = n\left(\frac{I}{4} - \frac{\pi^2 \hbar^2}{8m} n\right),$$

which is in close agreement with the DFT results shown in Figure 3. It leads to a critical doping density $n_c^{Stoner} = \frac{2mI}{\pi^2 \hbar^2}$, below which the paramagnetic state becomes less favorable. It also predicts that the spin-polarization energy reaches maximum $\Delta E_{max}^{Stoner} = \frac{mI^2}{8\pi^2 \hbar^2}$ at the doping density $n_{max}^{Stoner} = \frac{mI}{\pi^2 \hbar^2}$. We have plotted $\Delta E_{max}$ and $n_{max}$ extracted from DFT calculation against the Stoner model predictions in Figures 4 (a) and (b), and the theoretical values are in good agreement with the *ab initio* results.

According to this Stoner model, the FM instability come from the enhanced effective mass and Stoner parameter. Therefore, we can classify the strength of magnetism in GNRs according to these two parameters. As shown in Figures 4 (c) and (d), both the effective mass and the Stoner parameter decreases with the width of the nanoribbon. The electronic structures of GNRs falls into different groups depending on their width [37, 38]. Among them, the (3N+1)-a11-GNR, (3N+3)-a22-GNR, and (2N+2)-z211-GNR exhibit higher effective mass among each edge type and are thus more prone to having magnetism. For GNRs within the same family, the effective mass is inversely proportional to the effective width of the ribbon. This is due to the quantum confinement of graphene's Dirac dispersion π band [39, 40]. It follows a fitting formula $m = a/(w - w_0)$, as shown by the dashed line in Figure 4 (c), where $w$ is the physical ribbon width and $w_0$ is a correction that is found to be positive.

Meanwhile, the Stoner parameter is mainly decided by the local DOS at the Fermi energy and inversely proportional to the width $w$. The Stoner parameter $I$ is given by the integral $I = \int \gamma^2(\mathbf{r}) v_{xc}[n(\mathbf{r})] d\mathbf{r}$, where $\gamma(\mathbf{r}) = \sum_n \delta(\varepsilon_F - \varepsilon_n)|\psi_n(\mathbf{r})|^2 / N_0$ is the local DOS at the Fermi energy (normalized by the total DOS), and $v_{xc}[n(\mathbf{r})] = \left[-\frac{d^2}{dm^2} n(\mathbf{r}) \varepsilon_{xc}[n(\mathbf{r}), m]\right]\Big|_{m \to 0}$ is the second derivative of the exchange-correlation energy with respect to the magnetization [41]. When the same electronic state is confined within a region of width $w$, the local DOS will increase as $\gamma(\mathbf{r}) \propto 1/w$, which leads to an increase of the Stoner parameter as $I \propto 1/w$. This relation agrees well with the DFT calculation, as shown by the dashed line in Figure 4 (d). Particularly, we note that when comparing the value per atom, $I$ is around a constant value 3.4 eV, which agrees with the Hubbard U term estimated for graphene nanoribbons [42, 43] and is over 6 times larger than that of bulk 3d transition metals like iron [35].

As a result, both the maximum spin-polarization energy $\Delta E_{max}$ and the corresponding planar doping density is proportional to $\frac{1}{w^2(w-w_0)}$. The specific values of $w_0$ and proportionality constants for different types of GNRs are summarized in the Supplementary Table I. In Figures 4 (e) and (f), we show the scaling of $\Delta E_{max}$ and $n_{max}^{2D}$ with the ribbon width for hole-doped (3N+1)-a11-GNR, (3N+3)-a22-GNR, and electron-doped (2N+2)-

z211-GNR. It shows the emergence of magnetism in different GNRs as the ribbon width is narrowed down to around 1nm due to the strong cubic scaling. It is worth noting that the quantum confinement enhancement of the Stoner parameter is general and presents in all 1D structures. Therefore, this general mechanism can potentially be used to realize magnetism in other nanostructures as well.

In summary, we have predicted the existence of iterant magnetism and half-metallicity in doped narrow GNRs within first-principle DFT calculation. The magnetism originates from the bulk electronic state of the ribbon and is independent of the specific edge structures. From the Stoner theory, the magnetism come from the enhanced effective mass and Stoner parameter due to quantum confinement and its strength scales with the ribbon width in an inverse cubic relation. Given the widely observed doping in nanostructures, this magnetism is helpful for understanding a broad range of controversial measurements. Our findings propose a new route for realizing edge-independent magnetism in graphene nanoribbon and show that quantum confinement can be a general mechanism for realizing metal-free magnetism in nanostructures.

**Supporting Information**

Supporting Information includes the details of first-principles calculations and the model parameters.

**Acknowledgement**

We are supported by the National Science Foundation (NSF) CAREER Grant No. DMR-1455346, NSF EFRI-2DARE-1542815, and the Air Force Office of Scientific Research (AFOSR) grant No. FA9550-17-1-0304. The computational resources have been provided by the Stampede of Teragrid at the Texas Advanced Computing Center (TACC) through XSEDE.

**Figures:**

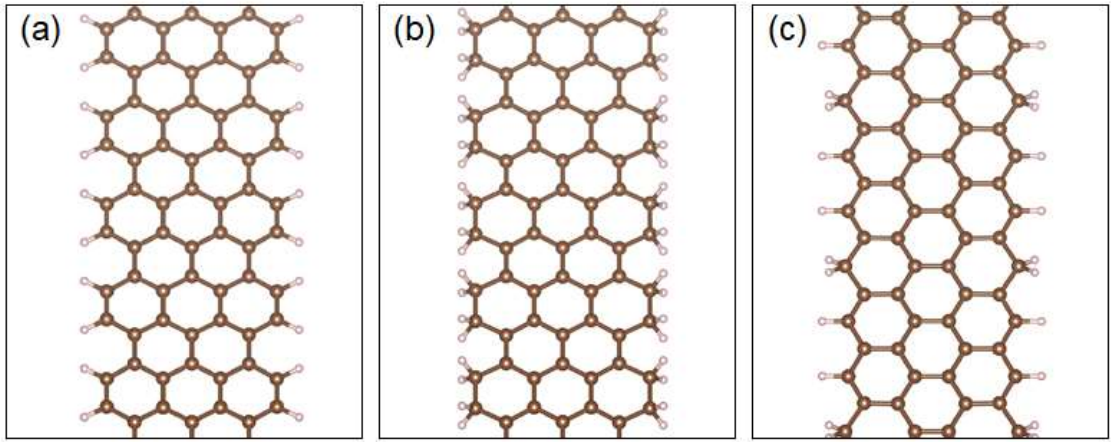

**Figure 1.** Examples of GNRs with three different type of edges considered in this work. From left to right are (a) 7-a11-GNR, (b) 7-a22-GNR and (c) 4-z211-GNR. The edge dangling bonds are passivated by hydrogen atoms.

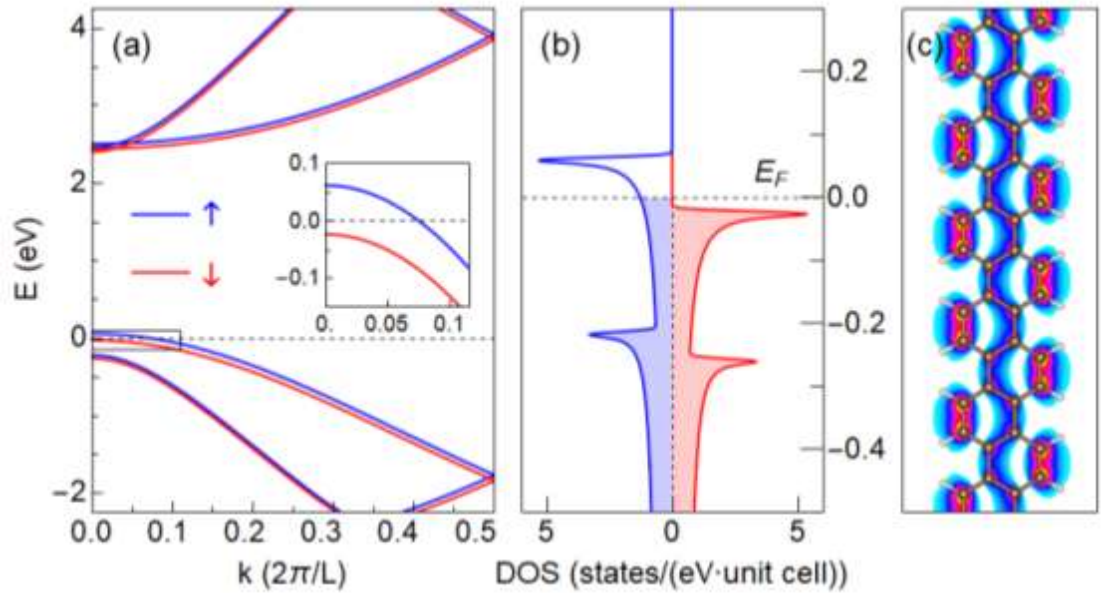

**Figure 2.** The DFT-calculated (a) band structure, (b) spin-projected DOS, and (c) real-space magnetization density of hole-doped 4-a11-GNR at hole density 0.35/nm. Fermi energy is set to be zero. Inset of (a) shows a zoomed-in view of band structure near the fermi energy, at the region indicated by the black rectangle on the main plot.

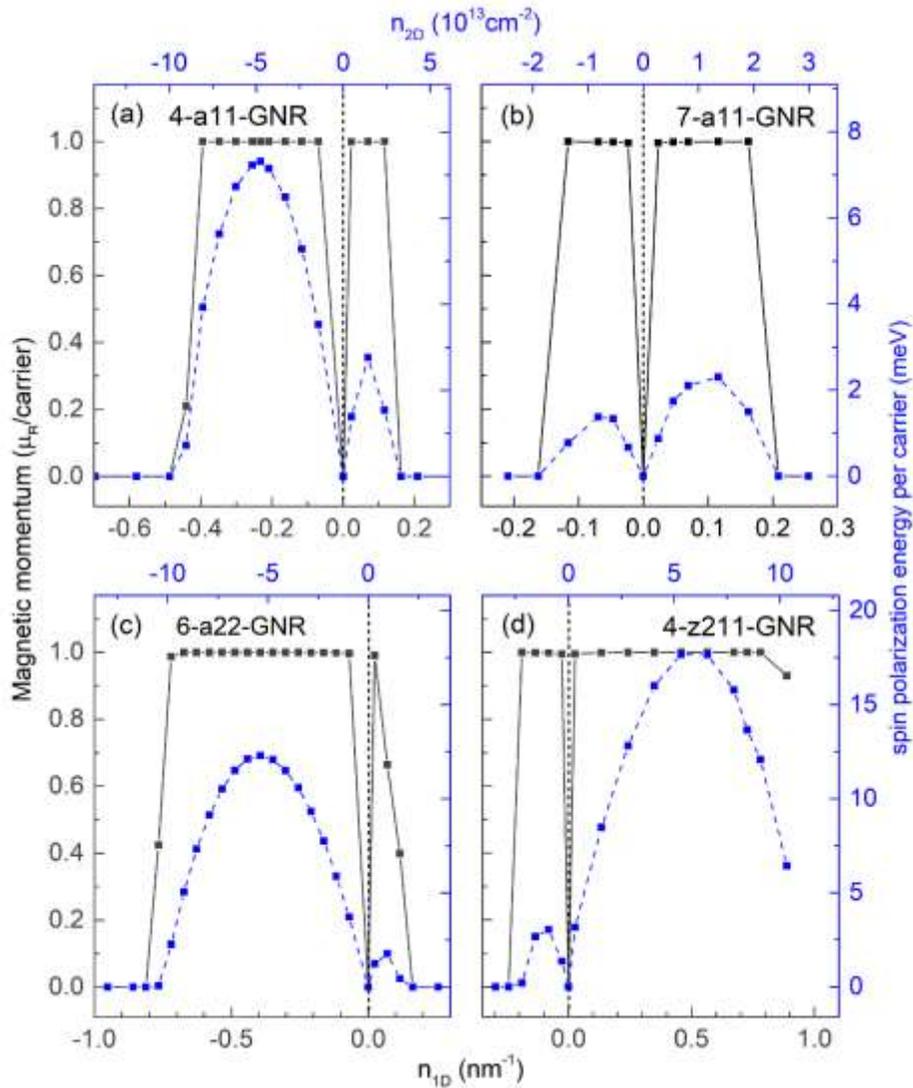

**Figure 3.** Magnetic momentum (black solid line) and spin polarization energy (blue dashed line) per carrier versus electron doping density for (a) 4-a11-GNR, (b) 7-a11-GNR, (c) 6-a22-GNR, and (d) 4-z211-GNR. The doping density is shown in 1D (number of electron or hole per nm) on the bottom axis and in 2D (number of electron or hole per $cm^2$) on the top axis. Positive and negative density correspond to electron and hole doping, respectively.

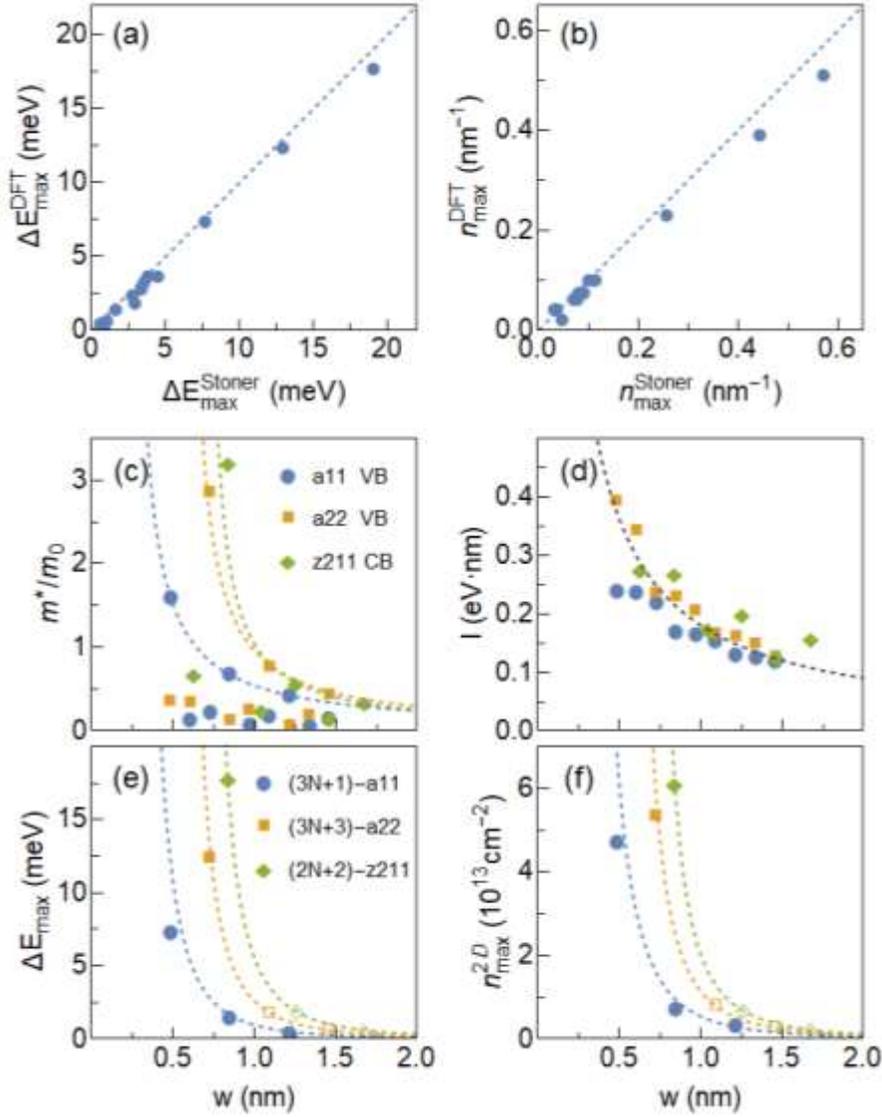

**Figure 4.** (a, b) Comparison between Stoner model prediction and DFT result for (a) the maximum spin-polarization energy and (b) the doping density corresponding to this maximum. (c, d) Evolution of the (c) effective mass and (d) Stoner parameter of the valence band with respect to the ribbon width, for the top valence band (VB) of a11- and a22-GNR and bottom conduction band of z211-GNR. The dashed lines on (c) are a fitting curve for (3N+1)-a11-GNR, (3N+3)-a22-GNR and (2N+2)-z211-GNR (N=1,2,3) according to the formula $a/(w - w_0)$. The dashed line on (d) is a fitting according to the formula $b/w$. (e, f) Predicted evolution of (e) the maximum spin-polarization energy and (f) the corresponding planar doping density with respect to the ribbon width, for hole-doped (3N+1)-a11-GNR, (3N+3)-a22-GNR and electron-doped (2N+2)-z211-GNR. The filled marks indicate the DFT results and the hollow marks indicate the predicted values.

# Supplementary Information: Edge-insensitive Magnetism and Half Metallicity in Graphene Nanoribbons

Shiyuan Gao [1] and Li Yang [1, 2] *

(1) Department of Physics, Washington University in St. Louis, St. Louis, Missouri 63130, USA

(2) Institute of Materials Science and Engineering, Washington University in St. Louis, St. Louis, Missouri 63130, USA

## 1. Computational Details

Our study is based on the ab initio pseudopotential projector-augmented wave DFT method [1] as implemented in the Vienna ab initio simulation package (VASP) [2]. The spin-polarized generalized gradient approximation with Perdew−Burke−Ernzerhof parametrization (GGA-PBE) [3] is used for the exchange-correlation functional. A cutoff energy of 800 eV for the plane-wave basis is used. Structural relaxation is performed with a converge criteria of $1\times10^{-2}$ eV/Å on force and a Γ-centered k-point grid of $1\times1\times20$. Electron self-consistency loop is performed with a converge criteria of $1\times10^{-8}$ eV for total energy and a Gaussian smearing of 0.0001 eV for electron occupation to ensure accurate convergence of the magnetic state. A Γ-centered k-point grid of $1\times1\times200$ is used for the armchair (a11 and a22) GNR and $1\times1\times100$ is used for the zigzag (z211) GNR. Rigid-band doping, which changes the total number of electrons in the unit cell with a compensating jellium background, is used to mimic the electrostatic doping [4-6].

## 2. The Magnetization density of different GNRs

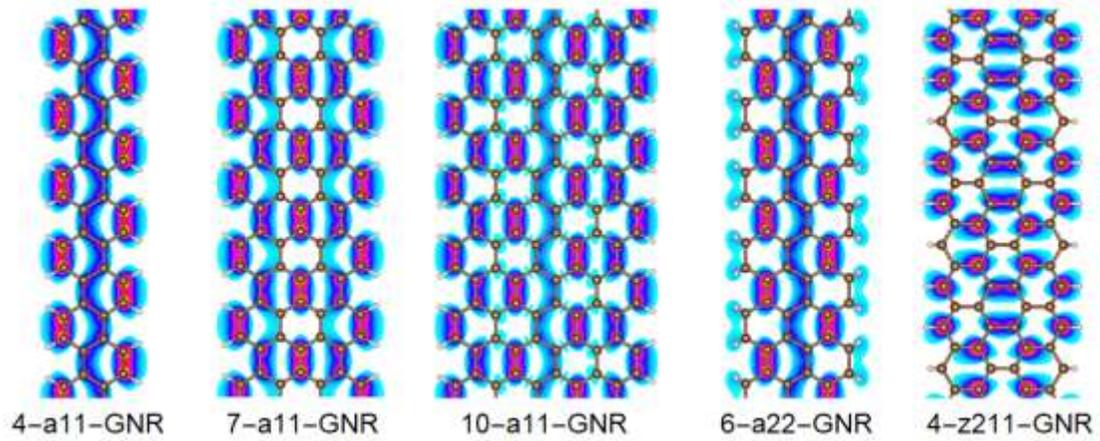

**Supplementary Figure 1**. Magnetization density plot for p-doped 4-a11-GNR, 7-a11-GNR, 10-a11-GNR, 6-a22-GNR, and 4-z211-GNR. The magnetization density does not decay when moving away from the edge, which shows the magnetism originate from the bulk state instead of edge state.

## 3. Magnetism and Half Metallicity in GNRs with edge defect

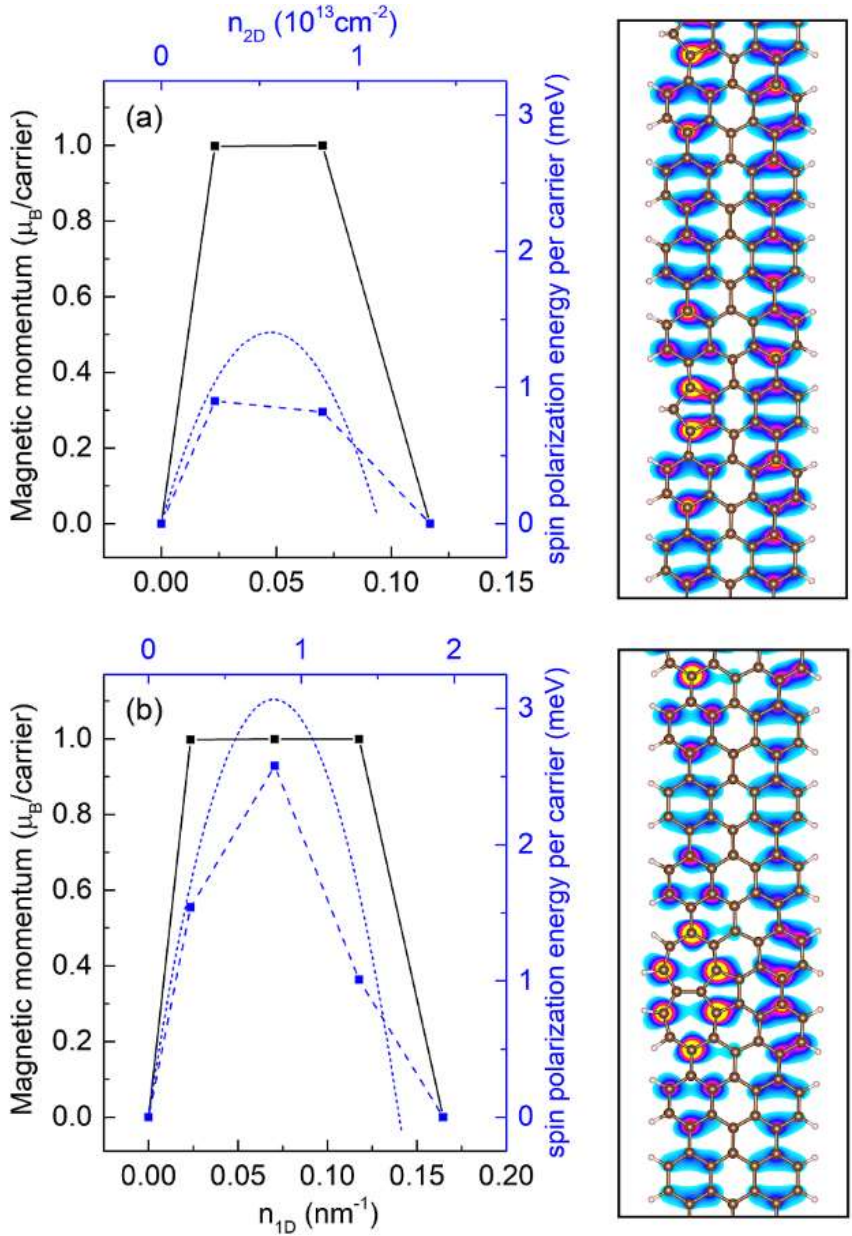

**Supplementary Figure 2.** Magnetic momentum and spin polarization energy per carrier versus hole doping density, and real-space magnetization density plot for 7-a11-GNR with two different types of defective edges. The first type (a) is a single pentagon due to the removal of one carbon atom. The second type (b) is a Stone–Wales defect. Both types of imperfections have been widely observed in experiments and can stand for general edge imperfections [7]. The defect density here is one per five unit cells, or 10% of the edge. The defect does hybridize with the bulk state and leads to different band effective mass and Stoner parameter. However, the doping-induced iterant ferromagnetism and half-metallicity remains, and the spin-polarization energy is close to that of pristine 7-a11-GNR. The blue dotted line shows the predicted spin-polarization energy from Stoner theory (Equation 1 of the main text). It agrees with the DFT result, showing that the same mechanism applies to these defective structures.

## 4. Fitting Parameters for the spin-polarization energy and planar doping density in different types of GNRs

| GNR type | $w_0$ (nm) | c (nm) | d (meV·nm$^3$) |
|---|---|---|---|
| n-doped (3N+1)-a11-GNR | 0.38 | 0.048 | 1.01 |
| p-doped (3N+1)-a11-GNR | 0.22 | 0.042 | 0.74 |
| n-doped (3N+3)-a22-GNR | 0.62 | 0.059 | 1.54 |
| p-doped (3N+3)-a22-GNR | 0.56 | 0.050 | 1.13 |
| n-doped (2N+2)-z211-GNR | 0.70 | 0.054 | 1.65 |
| p-doped (2N+2)-z211-GNR | 0.21 | 0.060 | 1.87 |
| n-doped (2N+1)-z211-GNR | 0.41 | 0.019 | 0.43 |
| p-doped (2N+1)-z211-GNR | 0.21 | 0.023 | 0.51 |

**Supplementary Table I**. Fitting Parameters for the maximum spin-polarization energy and corresponding planar doping density according to formula $n_{max}^{2D} \approx \frac{c}{w^2(w-w_0)}$ and $\Delta E_{max} \approx \frac{d}{w^2(w-w_0)}$. The ribbon width $w$ is counted starting half a C-C bond away from the outermost carbon atom.